\documentclass[a4paper]{article}
\usepackage{interspeech2009,amssymb,amsmath,epsfig}
\usepackage{graphicx}
\ninept
\def\reg{{\rm\ooalign{\hfil
     \raise.07ex\hbox{\scriptsize R}\hfil\crcr\mathhexbox20D}}}

\title{On the Mutual Information between Source and Filter Contributions for Voice Pathology Detection}

\makeatletter
\def\name#1{\gdef\@name{#1\\}}
\makeatother
\name{{\em Thomas Drugman, Thomas Dubuisson, Thierry Dutoit}}

\address{TCTS Lab, Facult\'e Polytechnique de Mons, Belgium \\
{\small \tt thomas.drugman@fpms.ac.be}}

\begin{document}
\maketitle

\begin{abstract}
This paper addresses the problem of automatic detection of voice pathologies directly from the speech signal. For this, we investigate the use of the glottal source estimation as a means to detect voice disorders. Three sets of features are proposed, depending on whether they are related to the speech or the glottal signal, or to prosody. The relevancy of these features is assessed through mutual information-based measures. This allows an intuitive interpretation in terms of discrimation power and redundancy between the features, independently of any subsequent classifier. It is discussed which characteristics are interestingly informative or complementary for detecting voice pathologies.
\end{abstract}


\section{Introduction}\label{sec:intro}

The acoustic evaluation of voice disorders is an essential tool for clinicians and is performed in a perceptive and objective way. On the one hand, the perceptive evaluation consists in  qualifying the voice disorder by listening to the production of a patient. This evaluation is performed by trained professionals who rate the phonation, using for instance the GRBAS scale \cite{GRBAS}. This approach suffers from the dependency on the experience of the listener and the inter- and intra-judges variability. On the other hand, the objective evaluation aims at qualifying and quantifying the voice disorder by acoustical, aerodynamic and physiological measures. Compared to methods based on electroglottography or high-speed imaging, the objective evaluation is cheaper, faster and more comfortable for the patient.

In order to provide objective tools to clinicians, a part of research in speech processing has focused on the detection of speech pathologies from audio recordings. Indeed it could be useful to detect disorders when perturbations are still weak, to prevent the degradation of the pathology, or to measure the voice quality before and after surgery in case of stronger disorders \cite{Gomez}.

Traditional methods of voice pathology detection generally rely on the computation of acoustic features extracted from the speech signal. From another point of view, video recordings of the vocal folds show that their behaviour is linked to the perception of different kinds of voice qualities, including pathologies (for instance, the incomplete closure of the folds can imply the perception of a \textit{breathy} voice \cite{Plumpe}). Isolating and parametrizing the glottal excitation should therefore lead to a better discrimination between normophonic and dysphonic voices. Such parametrizations of the glottal pulse have already been proposed both in time and frequency domains (\cite{Alku},\cite{Bozkurt}). Glottal source estimation and parametrization are for instance used to characterize different types of phonations \cite{Airas} or to derive biomechanical parameters for voice pathology detection \cite{Gomez}.


The goal of this paper is two-fold. First, a set of new features, mainly based on the glottal source estimation, is extracted through pitch-synchronous analysis. Secondly, we compare the proposed features according to their relevancy for the problem of automatic voice pathology detection. For this, we make use of information-theoretic measures. This is advantageous since the approach is independent of any classifier and allows an intuitive interpretation in terms of discrimination power and redundancy between the features.

The paper is structured as follows. In Section \ref{sec:FeatExtr}, the different features are described, some of them being extracted from an estimation of the glottal source obtained by the Iterative Adaptative Inverse Filtering (IAIF) method \cite{IAIF}. Section \ref{sec:MI} makes a review of interesting measures derived from the Information Theory and highlights their interpretation for a classification problem. Our experiments and results are detailed in Section \ref{sec:Exp}. It is discussed which features are particularly informative for detecting a voice disorder and which ones are interestingly complementary or synergic. Finally Section \ref{sec:conclu} concludes.

\section{Feature Extraction}\label{sec:FeatExtr}

The features considered in the present study characterize two signals. Some features are extracted from the speech signal, as in traditional methods of voice pathology detection. Others are extracted from the glottal source estimation in order to take into account the contribution of glottis in the production of a disordered voice. All proposed features are extracted from pitch-synchronous frames in voiced parts of speech. For this, the pitch and voicing decision are computed using the Snack library \cite{Snack} while Glottal Closure Instants (GCIs) are located on the speech signals using the DYPSA algorithm \cite{Dypsa}. In this work, glottal source-based frames (Section \ref{ssec:glottalFeat}) are two pitch-period long while others have a fixed length of 30 ms. In all cases, a GCI-centered Blackman window is applied. 


\subsection{Speech signal-based features}\label{ssec:speechFeat}

It is generally considered that the spectrum of the speech signal contains information about the presence of a pathology. The spectral content can be exploited using specific correlations between harmonics and formants although it must be noticed that these measures are dependent on the phonetic context \cite{Gomez}. Another way for summarizing the spectral content is to compute characteristics describing its repartition in energy. Most of the time these descriptors are designed to highlight the presence of noise.  For instance, it is proposed in \cite{Alonso} to divide the [0-11kHz] frequency range into 5 frequency bands and to compute the ratio of energy between some pairs of bands and between each band and the whole frequency range.

In the present study, it has been chosen to follow a similar idea as in \cite{Alonso} considering this time the perceptive mel scale. It is hoped this way to be close to human perception. For this, the power spectral density is weighted by a mel-filterbank consisting of 24 triangular filters equally spaced along the whole mel scale. The perceptive energy associated to each filter is computed as the sum of the weighted spectral density in the frequency range defined by this filter. Then three spectral balances are defined as:
\begin{equation} Bal1 = \frac{\sum_{i=1}^{4}{PE(i)}}{\sum_{i=1}^{24}{PE(i)}}\end{equation}
\begin{equation} Bal2 = \frac{\sum_{i=5}^{12}{PE(i)}}{\sum_{i=1}^{24}{PE(i)}}\end{equation}
\begin{equation} Bal3 = \frac{\sum_{i=13}^{24}{PE(i)}}{\sum_{i=1}^{24}{PE(i)}}\end{equation}
where $PE(i)$ denotes the perceptive energy computed in the $i^{th}$ mel band. The global repartition of spectral energy can also be captured in the spectral centroid, also known as spectral center of gravity ($CoG$). 

\begin{figure}[!ht]
  \centering
  \includegraphics[width=0.42\textwidth]{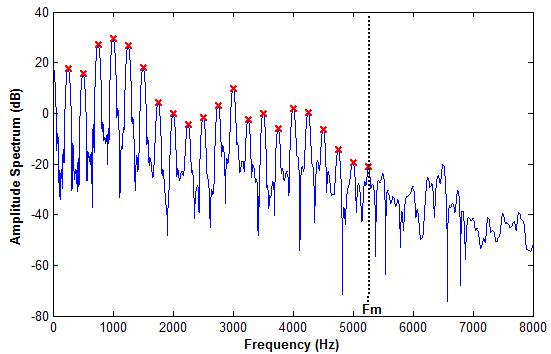}
  \caption{Example of maximum voiced frequency $Fm$ determination on a frame of normal voice. Harmonics are indicated (x) and $Fm$ corresponds to the last detected harmonic.}
  \label{fig:Fm}
\end{figure}

In addition, many studies attempt to quantify the presence of noise in the speech signal, as it is supposed to be linked to the perception of voice disorders. Many parameters have been proposed to quantify the importance of the spectral noise, one of the most popular being the Harmonic-to-Noise Ratio (HNR) \cite{Shama}. Basically this descriptor aims at quantifying the ratio between the energy of harmonic and noise components of the spectrum. HNR is here computed using the Praat software \cite{Praat} for comparison purpose. Besides we compute the so-called \emph{maximum voiced frequency} $Fm$, as suggested in the Harmonic plus Noise Model (HNM, \cite{HNM}). According to this model, the maximum voiced frequency demarcates the boundary between two distinct spectral bands, where respectively an harmonic and a stochastic modeling are supposed to hold. The higher $Fm$, the stronger the harmonicity, and consequently the weaker the presence of noise in speech. An example of determination of $Fm$ is exhibited in Figure \ref{fig:Fm}.

\vspace{-0.2cm}

\subsection{Glottal source-based features}\label{ssec:glottalFeat}

Most of methods estimating the glottal source rely on an inverse filtering step. Among these, a widely used algorithm is the Iterative Adaptive Inverse Filtering (IAIF, \cite{IAIF}) algorithm publicly available in the Aparat Toolkit \cite{Aparat}. The IAIF technique aims at iteratively estimating and removing the vocal tract contribution from the speech signal, so as to yield an approximation of the glottal source by inverse filtering. The amplitude spectrum for a voiced source generally presents a low-frequency resonance called \emph{glottal formant}, produced during the glottal open phase \cite{Bozkurt}. The glottal formant frequency ($Fg$) and bandwidth ($Bw$) are consequently two important characteristics of the glottal signal. As shown in \cite{Sigmap}, as long as the applied windowing is GCI-centered, relatively sharp and two-period long, considering only the left-part of the window makes a good approximation of the glottal open phase, allowing an accurate estimation of $Fg$ and $Bw$.

\begin{figure}[!ht]
  \centering
  \includegraphics[width=0.42\textwidth]{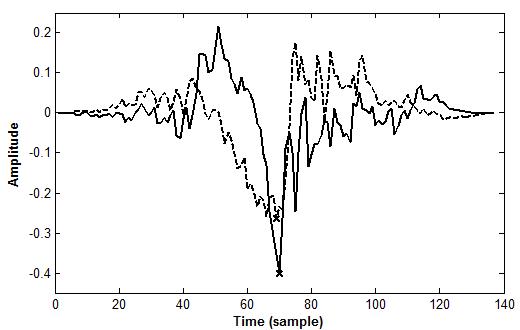}
  \caption{Comparison between a normal (solid) and pathological (dashed line) glottal frame. The minimum values around the GCI are indicated and are an image of the discontinuity strength occuring at this moment.}
  \label{fig:MinODGD}
\end{figure}

An example of glottal frames for both a normal and a pathological voices is exhibited in Figure \ref{fig:MinODGD}. It can be noticed that the discontinuity at the GCI is more significant for the normal voice. One possible way to quantify this is to take the minimum value at the GCI ($minGCI$) of energy-normalized frames. Apart from $Fg$, $Bw$ and $minGCI$, spectral balances $Bal1$, $Bal2$, $Bal3$ and center of gravity $CoG$ (as detailed in Section \ref{ssec:speechFeat}) are calculated on the glottal source frames.

\vspace{-0.2cm}

\subsection{Prosodic features}\label{ssec:prosoFeat}
It is often considered that dysphonic speakers have difficulty in maintaining stable prosodic characteristics during sustained vowels. These perturbations can be quantified by means of micro prosody measures \cite{Michaelis}. In this study, the prosodic features are inspired by these measures. Indeed, for each frame, $DeltaF0$ and $DeltaE$ are respectively defined as the variation of pitch and energy around their respective median value calculated over the whole phonation of sustained vowels (see Section \ref{sec:database}).

\section{Information-Theoretic Measures}\label{sec:MI}

The problem of automatic classification consists in finding a set of features $X_i$ such that the uncertainty on the determination of classes $C$ is reduced as much as possible \cite{FSBook}. For this, Information Theory \cite{Cover} allows to assess the relevance of features for a given classification problem, by making use of the following measures (where $p(.)$ denotes a probability density function):

\begin{itemize}
\item The entropy of classes $C$ is expressed as:
\begin{equation}
H(C)=-\sum_c{p(c)\log_2p(c)}
\label{eq:entropy}
\end{equation}
and can be interpreted as the amount of uncertainty on their determination.

\item The mutual information between one feature $X_i$ and classes $C$:
\begin{equation}
I(X_i;C)=\sum_{x_i}{\sum_c{p(x_i,c)\log_2\frac{p(x_i,c)}{p(x_i)p(c)}}}
\label{eq:MI}
\end{equation}
can be viewed as the information the feature $X_i$ conveys about the considered classification problem, i.e. the discrimination power of one individual feature.

\item The joint mutual information between two features $X_i$, $X_j$, and classes $C$ can be expressed as:
\begin{equation}
I(X_i,X_j;C)=I(X_i;C)+I(X_j;C)-I(X_i;X_j;C)
\label{eq:jointMI}
\end{equation}
and corresponds to the information that features $X_i$ and $X_j$, when \emph{used together}, bring to the classification problem. The last term can be written as:
\begin{equation}
\begin{split} I(X_i;&X_j;C)= \\
\sum_{x_i}\sum_{x_j}\sum_cp(x_i,x_j,c)\cdot&\log_2\frac{p(x_i,x_j)p(x_i,c)p(x_j,c)}{p(x_i,x_j,c)p(x_i)p(x_j)p(c)}
\end{split}
\label{eq:redundancy}
\end{equation}
An important remark has to be underlined about the sign of this term. It can be noticed from Equation \ref{eq:jointMI} that a positive value of $I(X_i;X_j;C)$ implies some \textbf{redundancy} between the features, while a negative value means that features present some \textbf{synergy} (depending on whether their association brings respectively less or more than the addition of their own individual information).

\end{itemize}

\section{Experiments}\label{sec:Exp}

\subsection{Database}\label{sec:database}
A popular database in the domain of speech pathologies is the MEEI Disordered Voice Database \cite{Kay}. This database contains sustained vowels and reading text samples, from 53 subjects with normal voice and 657 subjects with a large panel of pathologies. The recordings are linked to informations about the subjects (age, gender, smoking or not) and to analysis results. In this work, all the sustained vowels of the MEEI Database resampled at 16 kHz are considered.

\subsection{Mutual information computation}\label{ssec:MIcomput}

To evaluate the significance of the proposed features, the following measures are computed:

\begin{itemize}
\item the relative intrinsic information of one individual feature $\frac{I(X_i;C)}{H(C)}$, i.e. the percentage of relevant information conveyed by the feature $X_i$,
\item the relative redundancy between two features $\frac{I(X_i;X_j;C)}{H(C)}$, i.e. the percentage of their common relevant information,
\item the relative joint information of two features $\frac{I(X_i,X_j;C)}{H(C)}$, i.e. the percentage of relevant information they convey together.
\end{itemize}

For this, equations presented in Section \ref{sec:MI} are calculated. Probability density functions are estimated by a histogram approach. The number of bins is set to 50 for each feature dimension, which results in a trade-off between an adequately high number for an accurate estimation, while keeping sufficient samples per bin. Since features are extracted at the frame level, a total of 32000 and 107000 examples is available respectively for normal and pathological voices. Mutual information-based measures can then be considered as being accurately estimated. Class labels correspond to the presence or not of a dysphonia.

\subsection{Results}\label{ssec:results}

\begin{figure*}[!ht]
  \centering
  \includegraphics[width=1.00\textwidth]{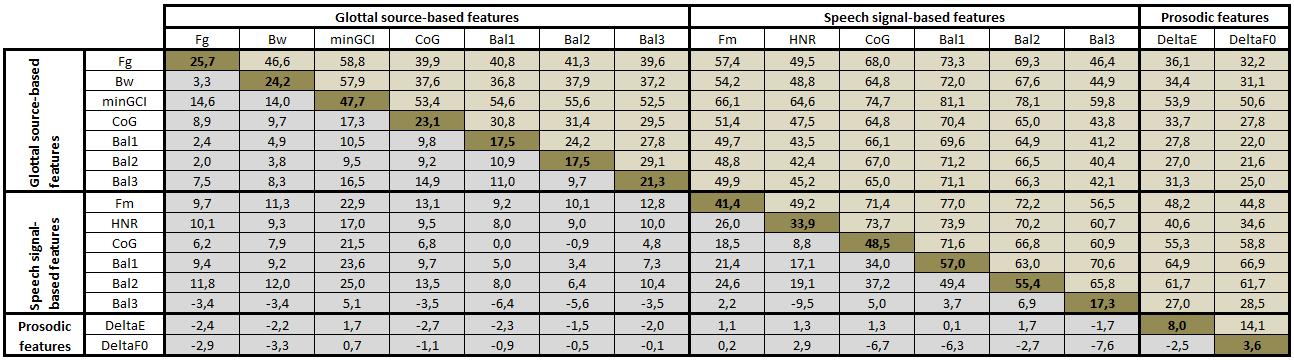}
  \caption{Mutual information-based measures for the proposed features. \emph{On the diagonal}: the relative intrinsic information. \emph{In the bottom-left part}: the relative redundancy between the two considered features. \emph{In the top-right part}: the relative joint information of the two considered features.}
  \label{fig:TabOfResults}
\end{figure*}

\begin{figure*}[!ht]
  \centering
  \includegraphics[width=0.90\textwidth]{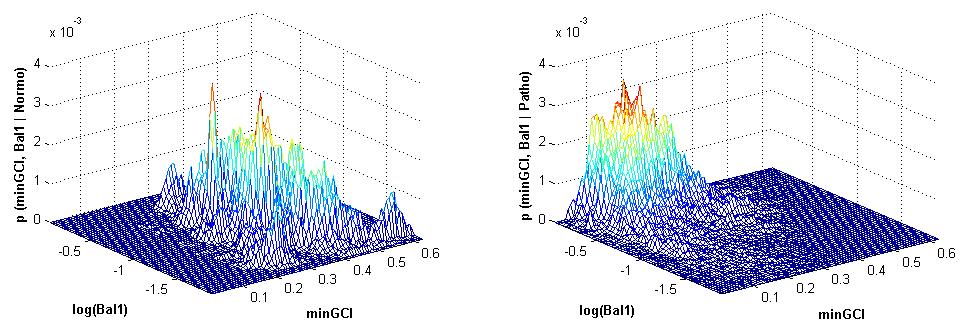}
  \caption{Example of separability for the two features giving the highest joint information.}
  \label{fig:ConditionalProba}
\end{figure*}

The values of the measures detailed in the previous section for the proposed features are presented in Figure \ref{fig:TabOfResults}. 
The diagonal indicates the percentage of relevant information conveyed by each feature. It turns out that features describing the speech spectrum contents are particularly informative, as well as $Fm$ and $minGCI$. The top-right part contains the values of relative joint information of two features, while the bottom-left part shows the relative redundancy between two features. Prosodic features convey few relevant information but are rather synergic with features from the two other categories. Although features describing the speech spectrum are intrinsically relevant, they are fairly redundant between them. More interestingly it can be noted that they are relatively complementary to glottal-based features. In particular the association of $minGCI$ and $Bal1$ brings the most important joint mutual information (81\%). Fig. \ref{fig:ConditionalProba} displays the joint distributions of these features for both normophonic and dysphonic voices. A clear separability between both classes can be observed. In addition it is confirmed that dysphonic speakers generally have diffculties in producing an abrupt glottal closure.

\section{Conclusion}\label{sec:conclu}

This paper focused on the problem of automatic detection of voice pathologies from the speech signal. Its goal was two-fold. First the use of the glottal source estimation was investigated and a set of new features was proposed. Second, extracted features were assessed through mutual information-based measures. This allowed their interpretation in terms of discrimination power and redundancy. It turned out that the spectral balances are particularly informative, as well as the maximum voiced frequency and the glottal discontinuity at the GCI. It was also shown that speech and glottal-based features are relatively complementary, while they present some synergy with prosodic characteristics. It is planned to extend the present work to the distinction between normophonic and dysphonic connected speech.

\section{Acknowledgments}\label{sec:Acknowledgments}

Thomas Drugman is supported by the ``Fonds National de la Recherche Scientifique'' (FNRS). The authors would like to thank the Walloon Region, Belgium, for its support (grant WALEO II ECLIPSE $\#$ 516009). This paper presents research results of the Belgian Network DYSCO, funded by the Interuniversity Attraction Poles Programme, initiated by the Belgian State, Science Policy Office. The scientific responsibility rests with its author(s).

\eightpt
\bibliographystyle{IEEEtran}

\end{document}